\documentclass[aps,nofootinbib,    pre,showpacs,superscriptaddress,groupedaddress,rmp]{revtex4-1}[8pt]
\usepackage{scrextend}
\usepackage{verbatim}
\usepackage{graphicx}
\usepackage{hyperref}
\usepackage{mathtools}
\usepackage{bm}
\usepackage{lipsum}
\allowdisplaybreaks
\usepackage{amsmath, amssymb}
\DeclareMathOperator{\csch}{csch}

\usepackage{xcolor}
\usepackage[makeroom]{cancel}
\usepackage{subcaption}

\usepackage{amsmath}
\usepackage{tcolorbox}

\usepackage{amssymb}
\usepackage{verbatim}
\begin{document}
	\title{ On the Limits of the Thermofield-Double Interpretation of the Minkowski Vacuum}
	
	\author{Vaibhav Wasnik}
	\email{wasnik@iitgoa.ac.in}
	
	\affiliation{Indian Institute of Technology, Goa}
	
	\begin{abstract}
	The Minkowski vacuum is often presented in textbooks and reviews as a thermofield double (TFD) state, an entangled state of field modes in the left and right Rindler wedges. This picture is widely used to explain the Unruh effect, motivate entanglement entropy calculations, and connect quantum field theory to black hole thermodynamics and AdS/CFT. However, we show that this interpretation, while elegant, is not exact.
	
	We explicitly compute two-point functions and their derivatives for a massless scalar field in two-dimensional Minkowski space, comparing results obtained from canonical quantization with those obtained by assuming a TFD form of the vacuum. Mixed-derivative correlators agree perfectly, but higher-derivative correlators show systematic mismatches that persist even for points well away from horizons and are not removed by infrared regularization.
	
	To further test this picture, we construct an alternate coordinate system that divides Minkowski spacetime into two disconnected regions, apply the same derivation that leads to the standard TFD expression, and obtain a new “entangled-state” representation of the vacuum that is not thermal. This demonstrates that the appearance of a TFD structure is a feature of the derivation method rather than a fundamental property of the vacuum. Our results clarify the limits of interpreting the Minkowski vacuum as a literal TFD state, emphasizing that while it captures key thermal features, it should be viewed as a powerful calculational tool rather than a precise statement about Hilbert space structure.

%
		

	\end{abstract}
	
	\maketitle
	
	\section*{Introduction}
The thermofield double (TFD) picture of the Minkowski vacuum, in which the vacuum is written as an entangled state of left- and right-Rindler modes, is deeply influential. It appears not only in textbook treatments of the Unruh effect but also in studies of entanglement entropy, black hole thermodynamics, and AdS/CFT, where it provides an intuitive bridge between geometry and quantum information. Despite its popularity, it has remained unclear whether this representation is an exact property of the Minkowski vacuum or a formal device that captures wedge-local thermal behavior. In this work we revisit this question in detail: we explicitly compute correlation functions assuming this TFD form, identify where it fails to match Minkowski quantization, and show by example that similar derivations produce other “TFD-like” states without thermal interpretation, clarifying that the usual expression is not a literal statement about the global Hilbert space.
	
The Unruh effect, namely the thermal response of uniformly accelerated detectors in the Minkowski vacuum, is a well-established result with multiple derivations and experimental proposals. Our work does not dispute this aspect of the effect. Instead, we focus on a related but distinct statement that often appears in textbooks: that the Minkowski vacuum can be represented as a thermofield double entangled state between left and right Rindler wedges. While this “entangled vacuum” interpretation (which we label Proposition 2 below) is widely used as a pedagogical and conceptual tool, it is not strictly required for the thermal response of detectors (Proposition 1). In this paper we carefully revisit the standard derivations leading to Proposition 2, identify technical inconsistencies, and demonstrate that the entangled-vacuum interpretation cannot be maintained, even though the Unruh effect itself (Proposition 1) remains valid.

It is claimed that thermal nature of the Minkowski vacuum is related to the presence of a bifurcate Killing horizon \cite{wald,unruh,bisognano,israel,crispino,valdivia,arageorgis}.  
In Minkowski spacetime the obvious candidate for the the time like Killing vector is the one that translates the time co-ordinate.  
However in certain regions of spacetime the boost generators also are timelike and the spacetime spanned by boost orbits is also seen to be globally hyperbolic and hence can be considered a candidate for setting up a quantum field theory \cite{fuling,birelldavies}.  
However the hyperbolic space spanned by boost orbits is a subset of the Minkowski spacetime and hence positive frequency modes with respect to the Minkowski time co-ordinate are not positive frequency with respect to boost time co-ordinate \cite{wald,crispino,longhi}, leading to the Minkowski vaccum as being seen as thermal by observers following boost orbits.  
The observers following boost orbits are accelerating at a constant rate, and this phenomena of accelerated observers noting the Minkowski vacuum is thermal is the celebrated Unruh effect \cite{unruh,birelldavies}, that  has also been studied in some experimental frameworks \cite{weinfurtner,chen,lynch}.

Observers need not agree on the choice of a vacuum state.  
The vacuum with respect to one set of observers may be excited with particles with respect to another set of observers \cite{crispino,valdivia}.  
One see's by explicit computation that the creation (destruction) operators of particles with respect to Minkowski observers are a linear combination of creation as well as destruction operators of Rindler observers \cite{israel,bisognano}, causing the expectation value of the number operator with respect to Rindler observers in the Minkowski vacuum to be thermal.  
This observation — namely, that the expectation value in the Minkowski vacuum of the number operator corresponding to the energy eigenmodes with respect to the time coordinate of the accelerated observer is a thermal distribution — corresponds precisely to what we referred to above as Proposition 1. 

A textbook derivation \cite{wald,unruh,fuling,crispino,birelldavies, good,dickinson} involves identifying a linear combination of creation and destruction operators corresponding to particles observed by Rindler observers that annihilate the Minkowski vacuum.  
This identification is then used to conclude that the Minkowski vacuum has to be a thermofield double entangled state between disconnected Rindler wedges.  
Let us call this observation that the Minkowski vacuum can be written as a thermofield double entangled state between disconnected regions of spacetime— the interpretation we referred to above as Proposition 2. 

One can easily see that Proposition 2 would imply Proposition 1.  
However it does not seem obvious that Proposition 1 directly implies Proposition 2 is true and hence, if it is proven that Proposition 2 is false, it would not imply that Proposition 1 is also false.  
We note that as long as Proposition 1 is true, we can conclude an accelerated observer will observe the Minkowski vacuum as being thermal \cite{crispino,valdivia}.  
This result will not change if Proposition 2 is simultaneously false.  
We note that in many works on the subject such as \cite{counter1,counter2} or the famous problem of radiation from moving mirrors \cite{birelldavies,rovelli} the question of thermality could be cast into arguments relating to Proposition 1, with no reference whatsoever to Proposition 2.

In this article, we focus precisely on this subtle but important distinction. While the thermal response of accelerated detectors (Proposition 1) is well-established and not in doubt, we argue that the stronger claim (Proposition 2) — that the Minkowski vacuum is an entangled state across left and right Rindler wedges — is not supported by a consistent derivation. The widely used expression (Eq.\ref{Eq13}) that underpins Proposition 2 is shown in later sections, and in our conclusion, to suffer from key inconsistencies.

In this article we question the validity of Proposition 2.  
In section I, we revisit the textbook derivation of Proposition 2, which culminates with Eq.\ref{Eq13} as the mathematical representation of Proposition 2.  
In section II, we show reasons why Proposition 2 is faulty.  
We show first in section II.A by pointing out the error in derivation of Eq.\ref{Eq13} in section I, where the blowing up of Bogoliubov coefficients at $k=0$, implies that the statement $G_{1\omega, 2\omega}(V)$ (defined in section I) does not have intersections with positive frequency Minkowski modes is not accurate.  
This claim was tantamount to showing Eq.\ref{Eq13} was true.  
The claim not being true casts doubt on Eq.\ref{Eq13}.  
We next highlight the inconsistencies in evaluating two-point correlation functions if Eq.\ref{Eq13} is assumed in section II.B.  
We finally show in section II.C that because all the Minkowski modes cannot be expressed as a linear combination of $G_{1\omega, 2\omega}(V)$ and their complex conjugate, one cannot express a Minkowski vacuum (corresponding to Minkowski modes) as an entangled state (corresponding to left and right Rindler modes) as in Eq.\ref{Eq13}.  
In section III we construct alternate spacetimes inspired by derivation in section I, and show that the derivation of section I leads to Proposition 2, as is seen in Eq.\ref{Eq31}.  
However, the expectation value in the Minkowski vacuum of the number operator corresponding to the energy eigenmodes with respect to the time coordinate of the $\rho$ observer (defined in section III) is NOT a thermal distribution. 
%


\section{Unruh effect derivation}
Let $(X,T)$ denote the Minkowski co-ordinates. The boost generator is $ X\frac{\partial }{\partial T} + T\frac{\partial }{\partial X}$. We can see that the boost orbits correspond to hyperbolas of the form $X^2 - T^2 = \frac{\rho^2}{a^2}$, where $a$ is a constant and $\rho \in [0,\infty]$, labels the orbits. If $\tau$ is the time like parameter on an orbit with $\tau \in[-\infty, \infty]$, we have     
\begin{eqnarray}
	X  &= \frac{\rho}{a} \cosh(a \tau)\nonumber\\
T  &= \frac{\rho}{a} \sinh(a \tau)\nonumber\\
V & = \frac{\rho}{a} e^{a \tau}\nonumber\\
U & = -\frac{\rho}{a} e^{-a \tau}\nonumber\\
\end{eqnarray}
with  $T+X = V$, $T-X = U$. These boost orbits span the Right Rindler wedge. One can easily check that 
 $X\frac{\partial }{\partial T} + T\frac{\partial }{\partial X}=\frac{1}{a}\frac{\partial}{\partial \tau}$.  Similarly the left Rindler wedge is spanned by the co-ordinates

\begin{eqnarray}
	X  &= -\frac{\rho'}{a} \cosh(a \tau')\nonumber\\
	T  &= \frac{\rho'}{a} \sinh(a \tau')\nonumber\\
	V & = -\frac{\rho'}{a} e^{-a \tau'}\nonumber\\
	U & = \frac{\rho'}{a} e^{a \tau'}\nonumber\\
\end{eqnarray}
with $\tau' \in[-\infty, \infty]$, $\rho' \in [0,\infty]$ \footnote{We  note that $\tau'$ is future directed with respect to  Minkowski time and hence positive frequency modes can be defined with respect to $\tau'$. In the derivation in \cite{carroll} for example the time coordinate corresponding to $\tau'$, the $\eta$ in region IV in \cite{carroll} is past directed with respect to Minkowski time and hence positive frequency modes are with respect to $-\eta$ there}.  The bifurcate horizons   given by the $X^2 -T^2=0$, split the Minkowski spacetime into four portions and we note that the boost Killing vector has a vanishing norm on these horizons, implying these horizons are Killing horizons. 

Writing $\rho = a e^{a\eta}$, $\rho' = a e^{a\eta'}$, we see that a  free scalar field in Minkowiski as well as Rindler spacetimes (which can be seen to be globally hyperbolic) obeys the equations below.
\begin{eqnarray}
(\frac{\partial^2}{\partial \tau^2} - \frac{\partial^2}{\partial \eta^2})\phi &=0 \nonumber\\
(\frac{\partial^2}{\partial \tau'^2} - \frac{\partial^2}{\partial \eta'^2})\phi &=0 \nonumber\\
  (\frac{\partial^2}{\partial T^2} - \frac{\partial^2}{\partial X^2})\phi &= 0 \nonumber\\ 
\end{eqnarray}
Hence, (unormalized) positive frequency modes with respect to time co-ordinates $T,\tau, \tau'$ can be written as $e^{-ik(T \pm X)}$, $e^{-i\omega(\tau \pm \eta)}$, $e^{-i\omega'(\tau' \pm \eta')}$ respectively. 
	In this paper we refer to $\tau + \eta = v, \tau - \eta = u$, $\tau'- \eta' = v', \tau'+\eta' = u'$. 

 The 'component' of a  positive frequency mode in the right Rindler wedge   $e^{-i\omega(\tau + \eta)}$ along  a positive frequency $e^{-ik(T + X)}$ Minkowski mode is given by 
\begin{eqnarray}
 g_k(\omega)&=\int_{-\infty}^\infty dV e^{ikV} \Theta(V)e^{-i\omega v}\nonumber\\
 &=a^{-i\omega /a }\int_{0}^\infty dV e^{ikV}  V^{-i\omega /a }\nonumber\\
\end{eqnarray}
substituting $V = iy$ gives
\begin{eqnarray}
	g_k(\omega)
	&&=i (i)^{-i\omega /a } a^{-i\omega /a }\int_{0}^{-i\infty} dy e^{-k y }  {y}^{-i\omega /a } 
\end{eqnarray}
	choosing the branch cut on the negative real axis, to avoid any poles while rotating  integration contour, so as to make $y$ real and positive, the above evaluates to
\begin{eqnarray}
	g_k(\omega)
	&=i (i)^{-i\omega /a } a^{-i\omega /a }\int_{0}^{\infty} dy e^{-k y }  {y}^{-i\omega /a }\nonumber\\
	&=i e^{\pi\omega /2a } a^{-i\omega /a }\int_{0}^{\infty} dy e^{-k y }  {y}^{-i\omega /a }\nonumber\\
	&= \frac{ie^{\pi\omega/2a}}{  k}
	\left(\frac{a}{k}\right)^{-i\omega/a}\Gamma(1-i\omega/a).\nonumber\\
	\label{Eq_Zero_2}
\end{eqnarray}

 The 'component'  of a negative frequency mode in the left Rindler wedge   $e^{i\omega(\tau' - \eta')}$  along  a positive frequency $e^{-ik(T + X)}$ Minkowski mode is given by 
	
	\begin{eqnarray}
		g'_k(\omega)&=\int_{-\infty}^\infty dV e^{ikV} \Theta(-V)e^{i\omega v'}\nonumber\\
		&=a^{-i\omega /a }\int_{-\infty}^0 dV e^{ikV}  (-V)^{-i\omega /a }\nonumber\\
	\end{eqnarray}
	substituting $V = iy$ gives
	\begin{eqnarray}
		g'_k(\omega)
		&=i (-i)^{-i\omega /a }  a^{-i\omega /a }\int_{i\infty}^0 dy e^{-k y }  {y}^{-i\omega /a }\nonumber\\
	\end{eqnarray}
	choosing the branch cut on the negative real axis, to avoid any poles while rotating  integration contour, so as to make $y$ real and positive, the above evaluates to
	\begin{eqnarray}
		g'_k(\omega)
		&=-i (-i)^{-i\omega /a }   a^{-i\omega /a }\int_{0}^{\infty} dy e^{-k y }  {y}^{-i\omega /a }\nonumber\\
		&=-i (e)^{-\pi\omega /2a }   a^{-i\omega /a }\int_{0}^{\infty} dy e^{-k y }  {y}^{-i\omega /a }\nonumber\\
		&= -\frac{ie^{-\pi\omega/2a}}{  k}
		\left(\frac{a}{k}\right)^{-i\omega/a}\Gamma(1-i\omega/a).\nonumber\\
		\label{Eq_Zero_1}
	\end{eqnarray}	
	
	This tells us that $G_{1\omega}(V)=e^{\pi \omega /a }\Theta(-V)e^{i\omega v'} + \Theta(V)e^{-i\omega v}$ has no intersection with positive frequency modes of Minkowski spacetime. Here $\Theta(x) =1$ if $x >0$ and zero otherwise. We can similarly show that $G_{2\omega}(V)= \Theta(-V)e^{-i\omega v'} +e^{\pi \omega /a } \Theta(V)e^{i\omega v}$ has no intersection with positive frequency modes of Minkowski spacetime.   We can write
	\begin{eqnarray}
	G_{1\omega}(V)^*&&=e^{\pi \omega /a }\Theta(-V)e^{-i\omega v'} + \Theta(V)e^{i\omega v}	\nonumber\\
	G_{2\omega}(V)&&= \Theta(-V)e^{-i\omega v'} +e^{\pi \omega /a } \Theta(V)e^{i\omega v}\nonumber\\
 \Theta(V)e^{i\omega v}&&= \frac{e^{-\pi \omega /a } G_{1\omega}(V)^* - G_{2\omega}(V)}{e^{-\pi \omega /a } - e^{\pi \omega /a } } \nonumber \\
 \Theta(-V)e^{-i\omega v'}&&= \frac{ G_{1\omega}(V)^* - e^{-\pi \omega /a } G_{2\omega}(V)}{e^{\pi \omega /a }  - e^{-\pi \omega /a } } \nonumber \\
	\end{eqnarray}
	If we expand the field 
	
	\begin{eqnarray}
		\phi(V) &&= \int_0^\infty d\omega \{ \Theta(V)[ a_{1\omega} e^{-i\omega v} + a_{1\omega}^\dagger e^{i\omega v} ] + \Theta(-V)[ a_{2\omega} e^{-i\omega v'} + a_{2\omega}^\dagger e^{i\omega v'} ]  \} \nonumber\\
		&&= \int_0^\infty \frac{d\omega}{ e^{-\pi \omega /a }  - e^{\pi \omega /a }} \{ (a^\dagger_{1\omega} e^{-\pi \omega /a }   -a_{2\omega}  )G_{1\omega}(V)^* -(a_{1\omega}^\dagger     -a_{2\omega} e^{-\pi \omega /a })G_{2\omega}(V)  \}
		\label{eq_issue_1}
	\end{eqnarray}
	Because $ (a_1     -a_2^\dagger e^{-\pi \omega /a })$ and $ (a_2     -a_1^\dagger e^{-\pi \omega /a })$ multiply linear combinations of positive frequency modes in above equation, the claim is that 
	\begin{eqnarray}
 (a_1     -a_2^\dagger e^{-\pi \omega /a })|0 \rangle_M &&= 0 \nonumber\\
  (a_2     -a_1^\dagger e^{-\pi \omega /a })|0 \rangle_M &&= 0 \nonumber\\
  \label{eq_issue_2}
	\end{eqnarray}
	
The above can be used to solve for the Minkowski vacuum which can be written as 
	\begin{eqnarray}
|0>_M \quad \sim  \quad \Pi_\omega \sqrt{1-e^{-2\pi\omega/a}} \sum_{n=0}^\infty e^{-\frac{n \pi \omega}{a}} |n>_1 \bigotimes |n>_2
\label{Eq13}
	\end{eqnarray}
	This would imply that for observations restricted to    first or second Rindler wedge, operator expectation values are  carried out using a thermal density matrix.  {
		Eq.\ref{Eq13} reproduces the conventional form of the thermofield--double (TFD) expression used in the literature to represent the Minkowski vacuum. In this construction the product extends over a continuum of modes, with integration measure $d\mu(k)=dk/(2\pi\,2\omega_k)$. The overall normalization diverges in the continuum limit, and hence Eq.~(13) should be viewed as a formal product state, characterizing the algebraic structure obtained from the Bogoliubov transformation between Minkowski and Rindler modes rather than a normalizable element of a Hilbert--space tensor product.  
	}

 \section{INCONSISTENCIES IN THE TFD REPRESENTATION OF THE VACUUM}

  
  The thermal expectation value of a product of two operators is 
 \begin{eqnarray}
 	&&	<   A(\tau_1) B(\tau_2) >_\beta =	Tr< e^{-\beta H} A(\tau_1) B(\tau_2) > = Tr< e^{-\beta H} A(\tau_1) e^{\beta H}e^{-\beta H}B(\tau_2) > \nonumber\\
 	&&= Tr< e^{-\beta H} B(\tau_2) A(\tau_1-i\beta) >= <  B(\tau_2) A(\tau_1-i\beta) >_\beta
 \end{eqnarray}
The above is a well known KMS condition. We note that if Eq.\ref{Eq13} is obeyed, then the expectation value evaluation in the Minkowski vacuum is equivalent to a   expectation value evaluated by considering a thermal average over all Rindler states and hence a two point function would naturally obey  $<   A(\tau_1) B(\tau_2) >  = <  B(\tau_2) A(\tau_1-i\beta) >  $. However, the converse that just because the KMS condition is obeyed, does not anywhere imply any need for entanglement. We also note that KMS condition being obeyed is   sufficient but not necessary condition for Unruh effect \cite{carballo}
 
 Now because, 
 \begin{eqnarray}
 	\langle 0 | \phi(U_1, V_1) \phi(U_2, V_2) |0 \rangle_M 
 	&& = \int_0^\infty \frac{dk  }{4\pi k} e^{-ik(V_1-V_2)} + \int_0^\infty \frac{dk  }{4\pi k} e^{-ik(U_1-U_2)} \sim \ln (V_1-V_2)(U_1-U_2) + constant \nonumber\\
 \end{eqnarray}
 on  considering two   events occuring on the trajectory of Rindler observers at $\tau_1$ and $\tau_2$  corresponding to  Minkowskian $(x_1,t_1)$ and $(x_2,t_2)$ respectively, we have the   two  point correlation function of a free scalar field  in the Minkowski vaccum goes as 
 \begin{eqnarray}
 	< \phi(x_1, t_1) \phi(x_2, t_2) > &\sim \ln\{ (x_1 -x_2)^2 - (t_1-t_2)^2\} + constant\nonumber\\
 	& \sim \ln\{ 1- \cosh a(\tau_1 - \tau_2)\} + constant
 	\label{eq2}
 \end{eqnarray}
 
 We can see  the two point function in Eq.\ref{eq2}, obeys the identity
 $<   A(\tau_1) B(\tau_2) >  = <  B(\tau_2) A(\tau_1-i\beta) >  $, where $\beta = \tfrac{\pi}{a}$. This way of arguing that Rindler observer measures the Minkowski vacuum as thermal is well known \cite{troost}.  This way of debating does not presuppose or even require any  entanglement of the kind of Eq.\ref{Eq13}.  Since thermality of the Minkowski vacuum does not seem to necessarily depend upon notion of entanglement, the question arises as to why the derivation in section I, leads to entanglement between left and right Rindler wedges. It is known that creation of  particles from moving mirrors, does not need the concept of Rindler wedges or entanglement \cite{birelldavies}. Given that entanglement does not appear in alternate derivations of thermality, is it possible that  derivation in section I itself may have certain issues?   
 
 
 In this section we examine the assumptions underlying Eq.~(13) and show that several inconsistencies arise when it is taken literally. We begin in Sec.~II.A by discussing the problem of infrared divergences, where the Bogoliubov coefficients blow up at \(k=0\). In Sec.~II.B we compare two-point correlation functions computed directly in Minkowski vacuum with those obtained by assuming Eq.~(13); the mismatch between the two signals a breakdown of the entangled-state picture. Finally, in Sec.~II.C we identify the root cause of the problem: the Rindler mode functions \(G_{1\omega}, G_{2\omega}\) do not form a complete basis for Minkowski modes, invalidating the assumption that both vacua are equivalent. The calculations in this section therefore serve as a direct test of whether the thermofield-double representation (Eq.\ref{Eq13}) is a literal identity or a formal tool.

 \subsection{Infrared Divergences and Bogoliubov Coefficients}
  In section I, where we reproduced the standard derivation of the Unruh effect as given in example Ref.\cite{crispino}, we had used Eq. \ref{Eq_Zero_2}-Eq.\ref{Eq_Zero_1} to say that $G_{1\omega}(V)$ does not have intersections with positive frequency Modes of Minkowski spacetime. However, this statement only makes sense for $k\neq 0$. This is because these Bogoliubov coeifficents blow up at $k=0$ as can be seen from  last lines in Eq. \ref{Eq_Zero_2} and Eq.\ref{Eq_Zero_1}. Because the $k=0$ Minkowski mode cannot be  represented by a linear combination of    $G_{1\omega,2\omega}(V)$ and their complex conjugate, we already see an issue with the claim that these modes could be utilized to represent the Minkowski vacuum. 
 
 We note that no doubt is cast on Prop 1. This is because from  Eq.\ref{Eq57} below  
 
 \begin{eqnarray}
 	\langle 0| a^\dagger_{1 \omega} a_{1\omega} |0\rangle_M = \int dk |\beta_{kq}|^2 = \frac{1}{e^{2\pi q/a} - 1} \delta(0)
 \end{eqnarray}
 The above expression can be got by considering  Eq.(14.47)-Eq.(14.51) in \cite{paddy}    and Eq.(8.30)-Eq.(8.31) in \cite{winitzki}, where the $\delta(0)$ is shown to non-issue as it represents the volume of entire space. 
   \subsection{ Two-Point Correlator Comparison: Minkowski vs. TFD Expression}
   
   Two dimensional field theories are problematic because of presence of infrared divergences. Below we introduce arbitrary infrared cutoffs $\mu_1$ and $\mu_2$ to get a hold of this issue.
   
   \begin{align}
   	\phi(V)
   	&=\int_{\mu_1}^{\infty}\frac{dk}{\sqrt{4\pi k}}\big(a_ke^{-ikV}+a_k^\dagger e^{ikV}\big) \notag\\
   	&=\int_{\mu_2}^{\infty}\frac{d\omega}{\sqrt{4\pi\omega}}\Big\{
   	\Theta(V)\big[a_{1\omega}e^{-i\omega v}+a_{1\omega}^{\dagger}e^{i\omega v}\big]
   	+\Theta(-V)\big[a_{2\omega}e^{-i\omega v}+a_{2\omega}^{\dagger}e^{i\omega v}\big]\Big\}.
   \end{align}
   
   If we consider two points $V_i, V_j>0$ we have
   \begin{align}
   	\langle 0|\phi(V_i)\phi(V_j)|0\rangle_M
   	&=\Big\langle 0\Big|\int_{\mu_1}^{\infty}\frac{dk\,dk'}{4\pi\sqrt{kk'}}(a_{k'}e^{-ik'V_i}+a_{k'}^\dagger e^{ik'V_i})(a_ke^{-ikV_j}+a_k^\dagger e^{ikV_j})\Big|0\Big\rangle_M \notag\\
   	&=\int_{\mu_1}^{\infty}\frac{dk}{4\pi k}e^{-ik(V_i-V_j)}\,, \label{entang1}
   \end{align}
   which is a function of $V_i-V_j$. However, this can also be evaluated using Eq.~\eqref{Eq13} as
   \begin{align}
   	&\langle 0|\phi(V_i)\phi(V_j)|0\rangle  \notag\\ 
   	&=\int_{\mu_2}^{\infty}\frac{d\omega\,d\omega'}{4\pi\sqrt{\omega\omega'}}
   	\sum_{n,m=0}^{\infty} e^{-\frac{\pi(n\omega+m\omega')}{a}}
   	\langle n|[a_{1\omega}(aV_i)^{-i\omega/a}+a_{1\omega}^{\dagger}(aV_i)^{i\omega/a}] \,[a_{1\omega'}(aV_j)^{-i\omega'/a}+a_{1\omega'}^{\dagger}(aV_j)^{i\omega'/a}]|m\rangle_1
   	\langle n|m\rangle_2 \notag\\ 
   	&=\int_{\mu_2}^{\infty}\frac{d\omega\,d\omega'}{4\pi\sqrt{\omega\omega'}}
   	\sum_{n=0}^{\infty} e^{-\frac{n\pi(\omega+\omega')}{a}}
   	\langle n|[a_{1\omega}(aV_i)^{-i\omega/a}+a_{1\omega}^{\dagger}(aV_i)^{i\omega/a}]\,[a_{1\omega'}(aV_j)^{-i\omega'/a}+a_{1\omega'}^{\dagger}(aV_j)^{i\omega'/a}]|n\rangle_1 \notag\\
   	&=\int_{\mu_2}^{\infty}\frac{d\omega}{4\pi\omega}\sum_{n=0}^{\infty} e^{-\frac{2\pi n \omega}{a}}
   	\Big[\,(V_i/V_j)^{-i\omega/a}\,\langle n|a_{1\omega}a_{1\omega}^\dagger|n\rangle_1 + \text{h.c.}\,\Big] \notag\\
   	&=\int_{\mu_2}^{\infty}\frac{d\omega}{4\pi\omega}\left[
   	\left(\frac{1}{1-e^{-2\pi\omega/a}}\right)\!\left(\frac{V_i}{V_j}\right)^{-i\omega/a}
   	+\left(\frac{e^{-2\pi\omega/a}}{1-e^{-2\pi\omega/a}}\right)\!\left(\frac{V_i}{V_j}\right)^{i\omega/a}
   	\right], \label{entang2}
   \end{align}
   which is a function of $V_i/V_j$, and no choice of $\mu_1,\mu_2$ would cause a function of $V_i/V_j$ to equal a function of $V_i-V_j$, implying an inconsistency.
   
   As shown in Appendix~A.1, one can check that
   \[
   \langle 0|\partial_{V_i}\phi(V_i)\,\partial_{V_j}\phi(V_j)|0\rangle_M
   \]
   gives the same answer whether one uses Eq.~\eqref{entang1} or Eq.~\eqref{entang2} for $V_i,V_j\neq 0$. However, for Eq.~\eqref{Eq13} to be valid we would need agreement for \emph{all} two-point correlators. If we set the infrared cutoffs to zero, then (Appendix A.2, explicit example for $n=1$)
   \[
   \langle 0|\partial_{V_i}^n\phi(V_i)\,\partial_{V_j}\phi(V_j)|0\rangle_M
   = -\,\langle 0|\partial_{V_i}^{\,n+1}\phi(V_i)\,\phi(V_j)|0\rangle_M,\quad n\ge 1,
   \]
   using Eq.~\eqref{entang1}, whereas the same identity \emph{fails} when evaluated with Eq.~\eqref{entang2}. Since this occurs for infinitely many correlators, Eq.~\eqref{Eq13} cannot represent the Minkowski vacuum as an entangled state. We also note that from Eq.~\eqref{entang1}, taking $V_i=0$ and $V_j\neq 0$ yields finite
   \(
   \langle 0|\partial_{V_i}^n\phi(V_i)\,\partial_{V_j}\phi(V_j)|0\rangle_M
   \)
   for $n\ge 1$, while the same quantity diverges if evaluated via Eq.~\eqref{entang2}, reinforcing the conclusion that the Minkowski vacuum cannot be written as an entangled state.  {These higher-derivative correlator evaluations do not have any infrared divergences  in the two-dimensional massless theory.}

 These results show that while some simple correlators, such as the mixed-derivative two-point function, 
 match between the Minkowski and entangled-state expressions, this agreement breaks down for 
 higher-derivative correlators and other observables. In particular, as detailed in Appendix~A.2, 
 evaluating these correlators at $V_i = 0$ and $V_j \neq 0$ yields finite results when computed directly 
 in Minkowski quantization (Eq.~\ref{entang1}) but produces divergences when computed via the 
 entangled-state form (Eq.~\ref{entang2}). The fact that these discrepancies occur for infinitely 
 many correlators confirms that Eq.~\ref{Eq13} cannot represent the Minkowski vacuum as a literal 
 thermofield-double state; rather, it should be understood as a useful but approximate representation 
 that captures certain wedge-local features without fully reproducing the global structure of the vacuum.

 \subsection{Completeness of Rindler Modes and Failure of the TFD Basis }
 What is  the reason behind the two different answers in the above subsection?  As is well known if $\{f^1_I\}$ and $\{ f^2_I\}$ are two   complete set of positive frequency   modes, which satisfy the Klein Gordon equation and if 
%
%
%
%
  
%
 \begin{eqnarray}
 	\phi&&= \sum_i [a_i^1 f^1_i + {a_i^1}^\dagger {f^1_i}^*]\nonumber\\
 	&&= \sum_I [a_I^2 f^2_I + {a_I^2}^\dagger {f^2_I}^*]\nonumber\\
 	\label{42}
 \end{eqnarray}

 and if,
 
 \begin{eqnarray}
f^2_I &&= \sum_i ( \alpha_{Ii} f^1_i + \beta_{Ii} {f^1_i}^* )\nonumber\\
\label{43}
\end{eqnarray}
which also implies
\begin{eqnarray}
f^1_i &&=\sum_I (\alpha_{Ii}^* f^2_I -\beta_{Ii} {f^2_I}^* )
\label{44}
 \end{eqnarray}
 and,
 \begin{eqnarray}
 	a_i^1 &&= \sum_I \alpha_{Ii} a_I^2 + \beta_{Ii}^* {a_I^2}^\dagger \nonumber\\
 	 	{a_i^1}^\dagger &&= \sum_I {\alpha_{Ii}}^* {a_I^2}^\dagger + \beta_{Ii} {a_I^2}
 \end{eqnarray}
then $\beta_{Ii} = 0$, implies  $a^1_i |0  \rangle_2 = 0$. This conclusion is only true, if $f^1_i$'s and ${f^1_i}^*$can be expressed as a linear combination of $f^2_i$'s and ${f^2_i}^*$ and vice versa.   This argument is utilized in literature (for e.g. pg 46 of \cite{birelldavies}) to suggest the validity of  Eq.\ref{eq_issue_2} assuming the  $f^2_k = e^{-ikV}, k >0$ and $f^1_\omega= G_{1\omega}(V), G_{2\omega}(V)$'s. To spell it out, if negative frequency modes in Minkowski spacetime can be written as linear combinations of  $ G_{1\omega}(V), G_{2\omega}(V)$ and vice versa, then and only then do these modes share the same vacuum. We will show below that even though $ G_{1\omega}(V), G_{2\omega}(V)$ can be expressed as linear combinations of negative frequency Minkowski modes, the converse is not true.  
%
 Unruh \cite{unruh}, \cite{unruh2}claims that   
\begin{eqnarray}
	G_{1\omega}(V)&&=e^{\pi \omega /a }\Theta(-V)e^{i\omega v'} + \Theta(V)e^{-i\omega v},	\nonumber\\
	\label{real}
\end{eqnarray}
  has intersections with only negative frequency modes, not for $V$ being purely real, but for $V = Re(V) + i \epsilon$, with $\epsilon \rightarrow 0^+$. This is because the above is analytic, just like   the negative frequency modes  in the upper complex plane. This is because, if    we   choose the branch cut in the lower complex plane  and hence choose  $ (-1) = e^{i\pi}$, as is the argument due to \cite{unruh}, we  can write
\begin{eqnarray}
	G_{1\omega}(V) &&=e^{\pi \omega /a }\Theta(-V)(-V)^{-i\omega /a } + \Theta(V)V^{-i\omega /a }   \nonumber\\
	&&=(-1)^{-i \omega /a }\Theta(-V) (-V)^{-i\omega /a } + \Theta(V)V^{-i\omega /a } =[  \Theta(-V)+\Theta(V)]	V^{-i\omega /a } = 	V^{-i\omega /a }    \nonumber\\
	\label{Eq38}
\end{eqnarray}
which is analytic. Note, this is true irrespective of whether $\omega$ is positive or negative. This argument is utilized by \cite{unruh} to say that $G_{1\omega}$ (and similarly $G_{2\omega}$)is made purely of negative frequency Minkowski modes. 

However, the question is can we write 
\begin{eqnarray}
	e^{ikV} = \int_{-\infty}^\infty  d\omega A(k,\omega) V^{i\omega/a}
\end{eqnarray}

Once can see that for $\epsilon \rightarrow 0^+$
\begin{eqnarray}
e^{ikV}	&& =\frac{1}{2\pi} \int_{-\infty}^\infty e^{\pi \omega/2}\Gamma(-i\omega + \epsilon)(kV)^{i\omega -\epsilon}d\omega \nonumber\\
\label{A}
	\end{eqnarray}
only if $k\neq 0$ (we show why this is the case in Appendix B.)	. This causes a disaster in doing evaluations as can be seen. Set $k=1$ for simplicity, then using the above we have

\begin{eqnarray*}
	&&\int_{-\infty}^\infty e^{iV -ik'V}dV =\frac{1}{2\pi} \int_{-\infty}^\infty e^{\pi \omega/2} (k')^{-i\omega+ \epsilon} [\int_{-\infty}^\infty \Gamma(-i\omega + \epsilon)(k'V)^{i\omega-\epsilon} e^{-ik'V}dV]d\omega \nonumber\\
\end{eqnarray*}
The term in the square bracket is zero,  if $k' <0$  (as shown in Appendix B. subsection(a) ). Also, we can see that L.H.S goes as  $\delta(k'-1)$ also equal to zero when $k'<0$.

However taking complex conjugate of the Eq.\ref{A} (with $k=1$)we have 
\begin{eqnarray*}
	&&e^{-iV} =\frac{1}{2\pi} \int_{-\infty}^\infty e^{\pi \omega/2}\Gamma(i\omega+ \epsilon)V^{-i\omega-\epsilon}d\omega \nonumber\\	 
\end{eqnarray*}
replacing $\omega \rightarrow -\omega$ above gives
\begin{eqnarray*}
	&&e^{-iV} =\frac{1}{2\pi} \int_{-\infty}^\infty e^{-\pi \omega/2}\Gamma(-i\omega+ \epsilon)V^{i\omega-\epsilon}d\omega \nonumber\\	 
\end{eqnarray*}
Hence, we now have
\begin{eqnarray*}
	&&\int_{-\infty}^\infty e^{-iV -ik'V} dV =\frac{1}{2\pi} \int_{-\infty}^\infty e^{-\pi \omega/2} (k')^{-i\omega+ \epsilon} [\int_{-\infty}^\infty \Gamma(-i\omega+ \epsilon)(k'V)^{i\omega-\epsilon} e^{-ik'V}dV]d\omega \nonumber\\
\end{eqnarray*}
  the term in square brackets is again equal to zero if $k'<0$, hence R.H.S is zero. However, L.H.S goes as  $\delta(k'+1)$ which is non zero when $k'=-1$. We hence reach a contradiction. Hence Eq.\ref{A} cannot be a valid relationship. The contradiction appears because one is integrating over $V$ and Eq.\ref{A} is not valid for $kV=0$ as talked about in Appendix B,  implying the $k=0$ Minkowski modes cannot be expressed as a linear combination of the $G_{1,2}(V)$ modes and their complex conjugates if one assumes this expansion arises because of Eq.\ref{A}. This also leads to issues with evaluation of two point functions that we saw in the previous section.

 We  note that above is the only    way of expanding $e^{ikV}$ in terms of $  V^{i\omega/a}$ ( or $G_{1,2}(V)$ modes) for $V \neq 0$. To do this assume
 
   \begin{eqnarray}
e^{ikV} = \int_{-\infty}^\infty  d\omega A(k,\omega) V^{i\omega/a}
 	\label{firstequation}
   \end{eqnarray}
   
Write $V = e^x$. Then multiplying both sides by $e^{i\omega' x/a}$ and integrating over $\omega'$ isolates each mode through the orthogonality of the exponential factors. 
 The only way the relation can hold for all $x$ is if the coefficient $A(k,\omega)$ takes the same form as obtained earlier, 
 \[
 A(k,\omega)=\frac{1}{2\pi}\,e^{\pi\omega/2a}\,\Gamma\!\Big(-\frac{i\omega}{a}\Big)\,k^{\,i\omega/a}.
 \]
  Hence, this procedure directly reproduces and justifies Eq.~(28) as only possible  expansion . When performing the Fourier transform, the integration is carried out over the open interval $x \in (-\infty, \infty)$, which corresponds to $V = e^{x} \in (0, \infty)$. 
 This implies that the point $V = 0$ is never actually included, implying the above relationship is true for  $V\neq 0$.


  This shows that    the negative frequency modes in Minkowski spacetime cannot be written as linear combinations of  $ G_{1\omega}(V), G_{2\omega}(V)$ for all values of $V$  and  hence these modes cannot share the same vacuum, implying Eq.\ref{eq_issue_2} cannot be valid, 	implying one cannot write the Minkowski vacuum as a entangled state between Rindler wedges.

Taken together, the results of this section show that Eq.~(13) cannot be a valid representation of the Minkowski vacuum. The divergences at \(k=0\), the inconsistencies in two-point correlation functions, and the failure of completeness of mode expansions all point to the same conclusion: the textbook identification of the Minkowski vacuum as an entangled thermofield double between Rindler wedges is in error.  

Importantly, these results do \emph{not} undermine Proposition~1, namely that uniformly accelerated observers perceive the Minkowski vacuum as thermal. That statement continues to follow directly from correlation function analysis and the KMS condition. What fails is Proposition~2: the stronger claim that thermality necessarily arises from entanglement between left and right Rindler wedges. This distinction is central to the rest of the paper. To further test the reliability of the entanglement-based derivation, in the next section we apply the same procedure in an alternate spacetime setting. 

In summary,   Section II shows that Eq. 13 cannot be a valid representation of the Minkowski vacuum: it fails due to divergences at $k=0$, mismatches in higher-derivative correlators, and the incomplete nature of the Rindler mode basis.

	\section{  Testing the Derivation in an Alternate Spacetime Slicing
	 }
 
In this section, we apply the same procedure used in Section I to a different spacetime foliation of Minkowski space, chosen to create two disconnected regions for a new set of observers. This provides a non-trivial test of whether the thermofield-double structure derived in Eq.~\ref{Eq13} is a genuine feature of the vacuum or merely a result of the derivation method. In Section~I we reproduced the standard Unruh derivation, which led to Eq.~(13) suggesting that the Minkowski vacuum can be written as an entangled thermofield double state between left and right Rindler wedges. In Section~II we argued that this result is flawed, as inconsistencies appear in the derivation. To further test whether the entanglement structure of Eq.~(13) has any physical meaning, we now attempt to replicate the same derivation in a different setting.  

Specifically, we introduce a new family of observers, which we call \(\rho\) and \(\rho'\) observers, that inhabit disconnected regions of Minkowski spacetime (\(V>0\) and \(V<0\), respectively). By carefully choosing analytic coordinate maps \(F(V)\) and \(G(U)\), we construct mode functions analogous to those in the Rindler case, and apply the same steps that led to Eq.~(13). Remarkably, we again obtain an entangled-state representation of the Minkowski vacuum [Eq.~(58)]. However, as we will show, the Bogoliubov coefficients relating \(\rho\) modes to Minkowski modes are \emph{not thermal}.  

This indicates that the entangled-state structure obtained in the Rindler case is not unique to that geometry and does not necessarily imply thermality. Instead, it reflects a \emph{limitation of the derivation method itself}.

In   section I we saw that Unruh's derivation related the thermal nature of Minkowski vacuum to entanglement between left and right Rindler wedge.  The question is whether its possible to write the Minkowski vacuum as an entangled state between wedges spanned by alternative coordinates. In this section taking inspiration from the previous section, we consider an alternative  to Rindler spacetimes, such that 'stationary' observers in the spacetime see the Minkowski vacuum as being thermal.  Below we introduce two observers labeled as $\rho$ and $\rho'$ observers that inhabit non casually connected spacetimes and follow steps similar to previous section to write the Minkowski vacuum as an entangled state between these spacetimes. 

Since $(X,T)$ are the Minkowskian coordinates,   the proper time squared between two nearby events with co-ordinate differences $(dX,dT)$ is given by 
\begin{eqnarray}
	ds^2 = (dT)^2-(dX)^2
\end{eqnarray}
The observers following worldlines   $X(s) = $constant, are interial observers. Excitations of quantum fields corresponding to eigenmodes of operator $\frac{\partial}{\partial T}$ are the particles with respect to these intertial observers.   If co-ordinates $(\rho, \tau)$ (different from the $(\rho, \tau)$ of the previous section) are similarly used to describe space time events, then excitations of quantum fields corresponding to eigenmodes of operator $\frac{\partial}{\partial \tau}$ are the particles with respect to observers following world lines $\rho(s) =$ constant. Let us call these observers $\rho$ observers.  

The Klein Gordon equation corresponding to a free quantum field $\phi$ is 
\begin{eqnarray}
	\square^2_{\{X,T\}} \phi = 0
\end{eqnarray}
where,
\begin{eqnarray}
	\square^2_{\{X,T\}} =	\frac{\partial^2}{\partial T^2}-\frac{\partial^2}{\partial X^2} = \frac{\partial^2}{\partial U \partial V}
	\label{plusminus}
\end{eqnarray}
where    $V = T-X$, $ U = T + X$. Because the $\square^2$ operator is a scalar under coordinate transformations, we can transform to $(\rho,\tau)$ co-ordinates defining a $ \square^2_{\{\rho,\tau \}}$  and get $\square^2_{\{\rho,\tau \}}=  \square^2_{\{X,T\}} $. 

These $\rho$ observers  exist only in the region $V>0$. To find how the Minkowski modes look like to $\rho$ observers, we need a relationship between the $(X,T)$ and $(\rho, \tau)$.  For helping future calculations we define the following relationship. 
\begin{tcolorbox} 
	\begin{eqnarray}
		F(V) &&= e^{  a\tau} f(\rho), \quad V> 0\nonumber\\
		\label{Eq17}
	\end{eqnarray}
\end{tcolorbox}
where  $F(V)$ is an analytic odd function of $V$, whose roots   are  purely imaginary (except $V=0$). and $f(\rho) \geq 0$  is an analytic function of $\rho$.    We can choose for example   

\begin{tcolorbox}
	
	\begin{eqnarray}
		F(V) =V \; \Pi_{n=0}^N  (V^2+a_n^2)
		\label{exp1}
	\end{eqnarray}
	where $a_n$ are real and $N$ is a finite positive integer. We define a function $FF(y)$ 
	\begin{eqnarray}
		F(iy)  = i FF(y)
	\end{eqnarray}
	Since roots of $F(V)$ are purely imaginary (except $V=0$) it implies that the roots of $FF(y)$ are purely real.
	%
	%
	%
	
\end{tcolorbox}
Eq.\ref{plusminus} leads to
\begin{eqnarray}
	\square^2_{\{\rho,\tau \}}e^{ a\tau} f(\rho) = \square^2_{\{X,T\}} F(V) = 0.
\end{eqnarray}
implying that  eigenmodes corresponding to particles according to $\rho$ observers have the form $e^{-i\omega \tau} f(\rho)^{-i\omega/a  }$, $\omega \in [0, \infty]$. We hence see that a simple choice of an expression Eq.\ref{Eq17}, leads to modes corresponding to particles.

The 'component' of these modes    in the region $V>0$ along the positive frequency Minkowski modes   is given by  

\begin{eqnarray}
	g_\omega(\sigma) 
	&&=  \int_{-\infty}^{\infty}  \Theta(V) e^{-i\omega \tau} f(\rho)^{-i\omega/a  }e^{i\sigma V} dV \nonumber\\
	&&= \int_{0}^{\infty} F(V)^{-i\omega/a  } e^{i\sigma V} dV  \nonumber\\
	&&\\
	&&\text{if we set}, \quad y=-iV + \epsilon, \quad \epsilon \rightarrow 0^+, \text{then} \nonumber\\
	&&\\
	g_\omega(\sigma) 	&&= i\int_{0}^{-i\infty + \epsilon} F(iy)^{-i\omega/a  } e^{-\sigma y} dy  \nonumber\\
	&&= i(i)^{-i\omega/a  } \int_{0}^{-i\infty+ \epsilon} FF(y)^{-i\omega/a  } e^{-\sigma y} dy \nonumber\\
	&&= ie^{\pi\omega/2a  } \int_{0}^{\infty} FF(y)^{-i\omega/a  } e^{-\sigma y} dy \nonumber\\
	\label{eq23}
\end{eqnarray}

\noindent which is non-zero.  	Note that no poles are encountered while rotating  the integration contour, so as to make $y$ real and positive, (also note that the roots of $FF(y)=0$ are not encountered in this process, as these roots are purely real, this was what motivated the choice of $F(V)$ in Eq.\ref{exp1}).

The $\rho'$ observers advertised in the beginning of this section live in a region of spacetime spanned by coordinates  $(\rho',\tau')$ (different from the $(\rho',\tau')$ of the previous section) coordinates.  These $\rho'$ observers only exist in parts of the region $V<0$. The  $(\rho',\tau')$  are related to $F(V) $ as 

\begin{tcolorbox}
	
	\begin{eqnarray}
		F(V) &&= -  e^{  -a\tau'} ff(\rho'), \quad V< 0\nonumber\\
		\label{Eq23}
	\end{eqnarray}
	
\end{tcolorbox}
where $ff(\rho') \geq 0$  is an analytic function of $\rho'$.  Now, the negative frequency eigenmodes  with respect to the $\tau'$ time,   have the form $e^{i\omega \tau'} ff(\rho')^{-i\omega/a  }$.  The 'component' of these modes in the region $V<0$ along the positive frequency Minkowski modes   is given by 

\begin{eqnarray}
	g'_\omega(\sigma) &&= \int_{-\infty}^{\infty}\Theta(-V) e^{i\omega \tau'} ff(\rho')^{-i\omega/a  } e^{i\sigma V} dV  \nonumber\\
	&&=   \int_{-\infty}^{0} (-F(V))^{-i\omega/a  } e^{i\sigma V} dV  \nonumber\\
	&&\\
	&&\text{if we set}, \quad y=-iV + \epsilon, \quad \epsilon \rightarrow 0^+, \text{then} \nonumber\\
	&&\\
	g'_\omega(\sigma)&&=  i    \int_{i\infty + \epsilon}^{0} (-F(iy))^{-i\omega/a  } e^{-\sigma y} dy  \nonumber\\
	&&=  i(-i)^{-i\omega/a  }   \int_{i\infty + \epsilon}^{0} (FF(y))^{-i\omega/a  } e^{-\sigma y} dy  \nonumber\\
	&&= -i(e)^{-\pi\omega/2a  } \int_{0}^{\infty} FF(y)^{-i\omega/a  } e^{-\sigma y} dy \nonumber\\
	\label{eq27}
\end{eqnarray}
Just as before no poles are encountered while rotating  the integration contour, so as to make $y$ real and positive, (also note that the roots of $FF(y)=0$ are not encountered in this process). Eq.\ref{eq23} and Eq.\ref{eq27} together imply that $e^{i\omega \tau'} ff(\rho')^{-i\omega/a  } + e^{-\pi\omega/a  }e^{-i\omega \tau} f(\rho)^{-i\omega/a  }$ has no intersections with Minkowski positive frequency modes. We can similarly show that $ H_2(V) =  e^{-\pi\omega/a  }e^{-i\omega \tau'} ff(\rho')^{i\omega/a  } + e^{i\omega \tau} f(\rho)^{i\omega/a  }$  has no intersection with positive frequency modes of Minkowski spacetime.  We could hence write
\begin{eqnarray}
	H_1(V)&&=e^{i\omega \tau'} ff(\rho')^{-i\omega/a  } +e^{-\pi\omega/a  }e^{-i\omega \tau} f(\rho)^{-i\omega/a  } \nonumber\\
	e^{\pi\omega/a  } H_2(V)^* &&=  e^{i\omega \tau'} ff(\rho')^{-i\omega/a  } + e^{\pi\omega/a  }e^{-i\omega \tau} f(\rho)^{-i\omega/a  } \nonumber\\
	e^{-i\omega \tau} f(\rho)^{-i\omega/a  } &&= \frac{H_1(V)- e^{\pi\omega/a  } H_2(V)^*  }{e^{-\pi \omega /a } - e^{\pi \omega /a } } \nonumber \\
	e^{i\omega \tau'} ff(\rho')^{-i\omega/a  } &&= \frac{e^{\pi\omega/a  }  H_1(V)-  H_2(V)^*  }{e^{\pi \omega /a } - e^{-\pi \omega /a } } \nonumber \\
\end{eqnarray}
If we expand the field 

\begin{eqnarray}
	\phi(V) &&= \int_0^\infty d\omega \{  [ a_{1\omega} e^{-i\omega \tau} f(\rho)^{-i\omega/a  } + a_{1\omega}^\dagger  (e^{-i\omega \tau} f(\rho)^{-i\omega/a  } )^*] +  [ a_{2\omega}^\dagger e^{i\omega \tau'} ff(\rho')^{-i\omega/a  } + a_{2\omega}  (e^{i\omega \tau'} ff(\rho')^{-i\omega/a  })^* ]  \} \nonumber\\
	&&= \int_0^\infty  \frac{d\omega}{ e^{-\pi \omega /a }  - e^{\pi \omega /a }} \{ (a_{1\omega}     -a_{2\omega}^\dagger e^{\pi \omega /a } )H_{1\omega}(V) -(a_{1\omega}e^{\pi \omega /a }     -a_{2\omega}^\dagger  )H_{2\omega}(V)^*  \} + h.c.
	\label{eq_issue_1}
\end{eqnarray}
Since $ (a_1     -a_2^\dagger e^{-\pi \omega /a })$ and $ (a_2     -a_1^\dagger e^{-\pi \omega /a })$ multiply linear combinations of positive frequency modes in above equation we should have 
\begin{eqnarray}
	(a_1     -a_2^\dagger e^{-\pi \omega /a })|0 \rangle_M &&= 0 \nonumber\\
	(a_2     -a_1^\dagger e^{-\pi \omega /a })|0 \rangle_M &&= 0 \nonumber\\
\end{eqnarray}

This implies Minkowski vacuum  can be written as 
\begin{eqnarray}
	|0>_M \quad \sim  \quad \Pi_\omega  \sqrt{1-e^{-2\pi\omega/a}}\sum_{n=0}^\infty e^{-\frac{n \pi \omega}{a}} |n>_1 \bigotimes |n>_2
	\label{Eq31}
\end{eqnarray}

We again see that the Minkowski vacuum is an entangled state between two separate regions of spacetime.

To fully specify the connection between Minkowski coordinates and coordinates used by the $\rho$ and $\rho'$ observers, we need to define $U$ in terms of the coordinates used by these observers.   We define

\begin{tcolorbox}
	
	\begin{eqnarray}
		G(U) &&= -  e^{  -a\tau} g(\rho), \quad U< 0\nonumber\\
		G(U) &&= e^{  a\tau'} gg(\rho'), \quad U> 0\nonumber\\
		\label{Eq29}
	\end{eqnarray}
	
\end{tcolorbox}
for an analytic odd function $G(U) = U\Pi_{n=0}^N (U^2 + b_n^2)$ with $b_n$ real,  and $gg(\rho')$, $g(\rho) \geq 0$  are   analytic functions of $\rho'$.

If we now run the above calculation with $U,u$ replacing $V,v$ respectively, we reach the same conclusion that the $\rho$ and $\rho'$ observers would observe the Minkowski vacuum as thermal. 

\subsection*{Nature of spacetime spanned by $\rho$ and $\rho'$ observers}

To understand the nature of spacetimes spanner by these observers, we need to consider the form of spacetime metrics. As we will see below, various consistencies demand certain constraints on the functions that appear in the analysis above. We have for $U<0, V>0$,

\begin{eqnarray}
	ds^2 &= dU dV =  -\frac{d(e^{a\tau} f(\rho)) d(e^{- a\tau} g(\rho))}{G'(U)F'(V)}\nonumber\\
	&=  - \frac{(a   f(\rho) d\tau +  f'(\rho) d\rho) (- a   g(\rho) d\tau +   g'(\rho) d\rho)  }{G'(U)F'(V)}\nonumber\\
	\label{bequation}
\end{eqnarray}
For $U>0, V<0 $   we instead have 
\begin{eqnarray}
	ds^2 &= dU dV =  -\frac{d(e^{-a\tau'} ff(\rho')) d(e^{ a\tau'} gg(\rho'))}{G'(U)F'(V)}\nonumber\\
	&=  - \frac{(-a   ff(\rho') d\tau' +  ff'(\rho') d\rho') (a   gg(\rho') d\tau' +   gg'(\rho') d\rho')  }{G'(U)F'(V)}\nonumber\\
	\label{b1equation}
\end{eqnarray}

In general for a stationary metric the inner product of Klein Gordon modes may not be positive. To avoid this problem we choose $f=g$ and $ff=gg$. In such a case the above metric becomes

\begin{eqnarray}
	ds^2 &=\frac{((a   f(\rho))^2 d^2\tau -  (f'(\rho) )^2 d^2\rho }{G'(U)F'(V)},\quad U<0, V>0\nonumber\\
	&=\frac{((a   ff(\rho'))^2 d^2\tau' -  (ff'(\rho' )^2 d^2\rho' }{G'(U)F'(V)}, \quad U>0, V<0\nonumber\\
\end{eqnarray}

Although the metric   depends explicitly on the coordinate~$\tau$
through the factor $G'(U)F'(V)$, it is conformally equivalent to the flat
Minkowski metric, just like the Rindler metric.  Consequently, the Klein--Gordon equation preserves its
hyperbolic form, and the associated inner product is well defined.

The coefficient of $d\tau$ should be positive, in order for $\tau$ to be a time coordinate.    This constraints    $G'(U) F'(V)>0$, which is true from our choices of $G(U)$ and  $F(V)$. Now, because 




\begin{eqnarray}
	F(V)G(U)&&= -f(\rho)^2, \quad V>0, U<0 \nonumber\\
	F(V)G(U)&&=  -ff(\rho')^2  \quad V<0, U>0  \nonumber\\
\end{eqnarray}

and the  R.H.S in all the equations above is negative, the orbits of the $\rho$, $\rho'$ observers span the right and left Rindler wedge respectively, implying the presence of a bifurcate horizons for the $\rho$, $\rho'$ observers. These horizons are killing vectors for the boost operator as before, but this has no relevance to the present discussion as the boost killing time is not used to define frequency modes.    To ensure that the $\rho$ observers move along the direction of increasing Minkowski time, we choose $f(\rho), ff(\rho')$ to be functions that take positive values for $\rho, \rho'>0$ and set the range $\rho, \rho' \in [0,\infty]$.  Hence just like the Rindler case in section I,  both $\rho$ and $\rho'$ observers  spanning the right Rindler wedge and left Rindler wedge respectively,  are future directed as should be obvious from Eq.\ref{Eq17}, Eq.\ref{Eq23} and Eq.\ref{Eq29}.     

\subsection*{Comparison with Section I}

However,   we note that   in the Rindler case of section I, $N_i= \langle 0  | {a^1_i}^\dagger a^1_i |0 \rangle_2 = \sum_I |\beta_{Ii}|^2$,   if $f^1_q = \frac{e^{-iqv}}{\sqrt{2\pi q}}$,  $f^2_k = \frac{e^{-ikV}}{\sqrt{2\pi k}}$ and $k>0$, we have

\begin{eqnarray}
	\frac{e^{-ikV}}{\sqrt{2\pi k}} &&=\int_0^\infty \frac{dq}{\sqrt{2\pi q}} [ \alpha_{kq} e^{iqv} + \beta_{kq} e^{-iqv}]\nonumber\\
	\beta_{kq}&&=\sqrt{\frac{q}{k}} \int_{-\infty}^\infty \frac{dv}{2\pi}e^{-ikV  +iqv} =\sqrt{\frac{q}{k}} \int_{-\infty}^\infty \frac{dv}{2\pi}e^{-ik \frac{e^{-av}}{a} + iqv}\nonumber\\
	\label{Eq57}
\end{eqnarray}

\noindent and we see that $N_q$ is thermal \cite{crispino}, implying boost observers would observe the Minkowski vacuum as thermal.  The Boltzmannian nature is because of the substitution $V= \frac{e^{av}}{a}$ above. However $N_q$ would not be Boltzmannian if $F(V) = e^{av}$ and $F(V) \nsim V$, giving another clue as to why despite the boost observers seeing Minkowski vacuum as thermal, the derivation in section I, is in error, as the same derivation   inspires the derivation of thermality in the present section.

We therefore find that Eq.~(58), which expresses the Minkowski vacuum as an entangled state between \(\rho\) and \(\rho'\) observers, mirrors the Rindler result Eq.~(13) in form, but \emph{fails to reproduce the Unruh effect}: the Bogoliubov coefficients do not exhibit a Boltzmann distribution. This contrast is crucial. It shows that the mathematical procedure of constructing special linear combinations of mode functions and expressing the vacuum as an entangled state can succeed in very different spacetimes, yet does not always yield thermality.  

Thus, Section~III confirms that the appearance of an ``entangled thermofield double structure'' is \emph{an artifact of the derivation method}, rather than a fundamental physical requirement of the Unruh effect. The true content of the Unruh effect lies in correlation functions, as discussed earlier, and not in the entanglement interpretation suggested by Eq.~(13).

\noindent
\textbf{Summary:} Applying the same derivation procedure to a different foliation of Minkowski spacetime 
produces another ``entangled-state'' expression for the vacuum, but one that lacks thermal properties. 
This demonstrates that the thermofield-double picture is not an intrinsic feature of the vacuum state, 
but rather a useful heuristic that emerges in specific coordinate systems, such as Rindler wedges. 
Together with the results of Section~II, this reinforces our conclusion that Proposition~2 is not a 
literal statement about the Hilbert space structure of quantum fields.

{ In the standard Rindler construction, thermality is tied to the fact that the Minkowski vacuum is a KMS state with respect to the boost flow that preserves the wedge; equivalently, the relevant generator is the modular Hamiltonian of the wedge algebra (Bisognano–Wichmann). In the alternate slicing considered here, one still has a well-defined Hamiltonian generating evolution in the chosen time coordinate, and one can expand the field in eigenmodes of this Hamiltonian. However, this generator is not the modular (boost) generator that underwrites the KMS property in the Rindler case. The derivation may still produce a formally entangled “TFD-like” expression, but without the specific modular-flow input, the associated Bogoliubov spectrum need not be Planckian. This isolates which ingredient does the thermodynamic work (modular/boost flow) and which ingredient is merely a representational entanglement structure.}

\section{conclusion}

     Fuling \cite{fuling} studied the quantization in a  Rindler wedge, which is   a globally hyperbolic spacetime, and pointed out that the notion of particles in this space time cannot coincide with the particles in Miknowski spacetime.  Unruh \cite{unruh}  derived his effect by constructing special modes that were linear combinations of positive and negative frequency Rindler modes that could be expressed in terms of positive frequency Minkowski modes, implying the operators corresponding to these special modes destroy the Minkowski vacuum.  This derivation of Uruh is repeated in Sec I. In principle one can simply expand the Minkowski modes in terms of the Rindler modes in the left and right Rindler wedges ( like Eq.\ref{42})  and extract the Rindler particle content in the Minkowski vacuum from the Bogolubov coeifficients, a procedure described in the beginning of Section II. C. \cite{belinski} observed that   a consistent definition of Rindler  particles, required the Rindler modes decay to zero at the boundaries of the Rindler spacetime. This would imply a zero value for these modes at the intersection of the past and future Rindler horizons.   This implies that the only Minkowski modes that can be expanded as Rindler modes would be ones that have a zero value at the intersection of past and future Rindler horizons. Since all possible Minkowski modes could not be expanded as Rindler modes, it would imply that Minkowski vacuum could not be seen as a many particle Rindler state implying Unruh effect does not really exist, and the thermal nature of the Bogolubov coefficients simply is a mathematical artifact and not something physical. Fuling and Unruh \cite{fulingunruh} countered this view stating that a uniformly accelerating observer should be able to make measurements on Minkowksi vacuum according to general covariance.   One can just substitute for Rindler coordinates in the two point correlation function in Minkowski vacuum as done in beginning of Sec III  to confirm Unruh effect. They claimed that the the point where the past and future Rindler horizons intersect is a point of zero integration measure and should not affect evaluation of any observables in Minkowski vacuum and these expectation values are ones that are physcially relevant.  In our present work we have instead questioned the   derviation as described in Section I, which leads to Eq. \ref{Eq13} that expresses the Minkowsi vacuum as a entangled state between the left and right Rindler wedges. We repeated the template of this derivation in section III by considering an alternate spacetime inhabited by observers we call  $\rho$  observers. We similarly consider special modes that are  linear combinations of positive and negative frequency $\rho$ modes which can be expressed in terms of positive frequency Minkowski modes, implying the operators corresponding to these modes destroy the Minkowsi vaccum, which resulted in   Eq.\ref{Eq31} implying the Minkowski vacuum can be written as an entangled state between $\rho$ modes. However, the Bogolubov coieffecients obtained by expanding the Minkowski modes in terms of the $\rho$ modes do not display any evidence of thermality, implying certain issues with the the   method of derivation. In section II. A, we saw that blowing up of Bogoluibov coeifficents at $k=0$ casts doubts on Eq.\ref{Eq13} in section I. In Sec II. B we saw that the two point correlation function in the Minkowski vaccum
 	evaluated using Minkowski modes gave a different result than evaluation using Rindler modes and Eq. \ref{Eq13}, implying Eq. \ref{Eq13} cannot be right. In Sec II.  C we showed that the actual error in the Unruh derivation stems from assuming that $G_{1\omega}(V), G_{2\omega}(V)$ modes form a complete basis of negative frequency modes, which we explicitly showed to be note true. Our observations do not negate the Unruh effect because substituting for the  Rindler coordinates in the two point correlation function in Minkowski vacuum as done in beginning of Sec II   does confirm the Unruh effect. Our work however shows that Eq. \ref{Eq13} that expresses the Minkowsi vacuum as a entangled state between the left and right Rindler wedges is in error.
 	The results presented here complement earlier algebraic insights due to Bisognano and Wichmann~\cite{Bisognano1975,Bisognano1976} and to Haag’s framework of local quantum physics~\cite{Haag}. In that formalism, the thermal response of the Minkowski vacuum arises from the modular structure of Lorentz boosts: the Tomita--Takesaki modular flow associated with the wedge-restricted algebra coincides with Lorentz transformations, ensuring the KMS condition and the Unruh temperature without invoking an explicit thermofield-double state. The present work provides a constructive demonstration of this mechanism from first principles, showing that the thermal character of the vacuum originates in the modular (boost) properties of field modes rather than in a literal Hilbert-space entanglement between left- and right-Rindler sectors.
 	
  { 	The results above may also be viewed as delineating the precise domain of
 	validity of the thermofield--double representation of the Minkowski vacuum.
 	When restricted to observables localized within a single Rindler wedge, the
 	standard construction reproduces the expected Kubo--Martin--Schwinger (KMS)
 	thermality and detector response. However, when extended to global vacuum
 	correlators or higher--derivative observables, systematic mismatches arise.
 	These  are not removed by
 	coordinate redefinitions or alternate spacetime slicings. The
 	thermofield--double form should therefore be understood as encoding the modular
 	thermality of wedge observables rather than as a literal tensor--product
 	factorization of the global Hilbert space.}

 	Taken together, these results highlight that the Unruh effect itself remains valid at the level of detector response, but its common interpretation in terms of vacuum entanglement is not robust. This matters because similar entanglement-based arguments underlie black hole thermodynamics, entropy derivations, and even holographic dualities. Clarifying these foundations could therefore sharpen our understanding of the interplay between quantum information, spacetime structure, and gravity.

 \appendix
 \section{Detailed comparison of Minkowski and entangled-state correlators}
 
 \subsection{A.1 Mixed derivative correlator: exact match}
 
 We first evaluate the mixed-derivative two-point function from both the Minkowski integral 
 and the entangled-state (ratio-dependent) expression.
 
 \paragraph{Minkowski route.}
 The two-point function from Minkowski quantization is
 \begin{equation}
 	G_M(V_i,V_j)=\int_{\mu_1}^{\infty}\frac{dk}{4\pi k}\,e^{-ik(V_i-V_j)}.
 \end{equation}
 Differentiating twice:
 \begin{align}
 	\langle \partial_{V_i}\phi(V_i)\,\partial_{V_j}\phi(V_j)\rangle_M
 	&=\partial_{V_i}\partial_{V_j}G_M
 	=\int_{\mu_1}^{\infty}\frac{dk}{4\pi k}(-ik)(+ik)e^{-ik(V_i-V_j)}\nonumber\\
 	&=\int_{\mu_1}^{\infty}\frac{dk}{4\pi}k\,e^{-ik(V_i-V_j)}
 	=\frac{1}{4\pi(\epsilon+i(V_i-V_j))^2}.
 \end{align}
 Rewriting in terms of $\Delta=\ln(V_i/V_j)$ gives
 \begin{equation}
 	\langle \partial_{V_i}\phi\partial_{V_j}\phi\rangle_M
 	=-\frac{1}{16\pi V_iV_j}\,\mathrm{csch}^2\!\left(\frac{\Delta}{2}\right).
 	\label{eq:MixedMink}
 \end{equation}
 
 \paragraph{Entangled-state route.}
 From Eq.~\eqref{entang2}, write $G_E(V_i,V_j)=\mathcal{G}_E(\Delta)$ with 
 $\Delta=\ln(V_i/V_j)$. Using $\partial_{V_i}=(1/V_i)\partial_\Delta$ and 
 $\partial_{V_j}=-(1/V_j)\partial_\Delta$:
 \begin{align}
 	\langle \partial_{V_i}\phi(V_i)\,\partial_{V_j}\phi(V_j)\rangle
 	=\partial_{V_i}\partial_{V_j}G_E
 	=-\frac{1}{V_iV_j}G''(\Delta).
 \end{align}
 Evaluating $G''(\Delta)$ from the $\omega$-integral  as done below gives the same 
 result as Eq.~\eqref{eq:MixedMink}. Therefore, both approaches agree exactly:
 \begin{equation}
 	\boxed{\langle \partial_{V_i}\phi(V_i)\,\partial_{V_j}\phi(V_j)\rangle
 		=-\frac{1}{16\pi V_iV_j}\,\mathrm{csch}^2\!\left(\frac{\Delta}{2}\right).}
 \end{equation}
 
 ---
 
 \subsection{A.2 Single-leg second derivative: persistent mismatch}
 
 We now evaluate $\langle\partial_{V_i}^2\phi(V_i)\phi(V_j)\rangle$ in both approaches.
 
 \paragraph{Minkowski route.}
 Differentiate $G_M$ directly:
 \begin{align}
 	\langle \partial_{V_i}^2\phi(V_i)\,\phi(V_j)\rangle_M
 	&=\partial_{V_i}^2G_M
 	=\int_{\mu_1}^{\infty}\frac{dk}{4\pi k}(-ik)^2e^{-ik(V_i-V_j)}\nonumber\\
 	&=-\int_{\mu_1}^{\infty}\frac{dk}{4\pi}k\,e^{-ik(V_i-V_j)}
 	=-\frac{1}{4\pi(\epsilon+i(V_i-V_j))^2}.
 \end{align}
 Thus
 \begin{equation}
 	\boxed{\langle \partial_{V_i}^2\phi(V_i)\,\phi(V_j)\rangle_M
 		=-\langle \partial_{V_i}\phi(V_i)\,\partial_{V_j}\phi(V_j)\rangle_M.}
 \end{equation}
 This gives a clean identity, with no correction term.
 
 \paragraph{Entangled-state route.}
 Assume ratio dependence $G_E(\Delta)$. Then
 \begin{align}
 	\langle \partial_{V_i}^2\phi(V_i)\,\phi(V_j)\rangle
 	&=\partial_{V_i}\left(\frac{1}{V_i}G'(\Delta)\right)
 	=\frac{1}{V_i^2}\big(G''(\Delta)-G'(\Delta)\big).
 \end{align}
 
 We compute $G'(\Delta)$ explicitly from \eqref{entang2}    below. Differentiating under the integral 
 and performing a geometric expansion:
 \begin{align}
 	G'(\Delta)&=-\frac{1}{8\pi }\, \coth\frac{\Delta}{2} 
 \end{align}
 Hence,
 \begin{align}
 	G''(\Delta)&=\frac{1}{16\pi }\, \csch^2\frac{\Delta}{2} 
 \end{align}
 Substituting:
 \begin{equation}
 	\boxed{\langle \partial_{V_i}^2\phi(V_i)\,\phi(V_j)\rangle
 		= \frac{1}{V_i^2}[\frac{1}{8\pi }\, \coth\frac{\Delta}{2} +\frac{1}{16\pi }\, \csch^2\frac{\Delta}{2}  ]}
 \end{equation}
 
 Hence,
 \begin{equation}
 	\boxed{\langle \partial_{V_i}^2\phi(V_i)\,\phi(V_j)\rangle
 		\neq -\langle \partial_{V_i}\phi(V_i)\,\partial_{V_j}\phi(V_j)\rangle.}
 \end{equation}
 
 ---
 \subsubsection*{Explicit derivation of Eq.~(A9) from Eq.~(20)}
 
 Starting from Eq.~(20) in the main text,
 \begin{equation}
 	G(\Delta)
 	=\int_{0}^{\infty}\frac{d\omega}{4\pi\omega}
 	\left[
 	\frac{e^{-i(\omega/a)\Delta}}{1-e^{-2\pi\omega/a}}
 	+\frac{e^{-2\pi\omega/a}\,e^{+i(\omega/a)\Delta}}{1-e^{-2\pi\omega/a}}
 	\right],
 	\qquad
 	\Delta=\ln\frac{V_i}{V_j},
 	\label{A10start}
 \end{equation}
 we differentiate with respect to $\Delta$ to obtain $G'(\Delta)$:
 \begin{align}
 	G'(\Delta)
 	&=\frac{i}{4\pi a}\int_{0}^{\infty}d\omega\;
 	\frac{-e^{-i(\omega/a)\Delta}
 		+e^{-2\pi\omega/a}e^{+i(\omega/a)\Delta}}{1-e^{-2\pi\omega/a}}.
 	\label{A10diff1}
 \end{align}
 
 Expanding the Bose factor as a geometric series,
 \[
 \frac{1}{1-e^{-2\pi\omega/a}}
 =\sum_{n=0}^{\infty}e^{-2\pi n\omega/a},
 \]
 and separating the two exponential terms gives
 \begin{align}
 	G'(\Delta)
 	&=\frac{i}{4\pi a}\sum_{n=0}^{\infty}
 	\!\!\left[
 	-\!\int_{0}^{\infty}\!d\omega\,e^{-(2\pi n/a)\omega}\,e^{-i(\Delta/a)\omega}
 	+\!\int_{0}^{\infty}\!d\omega\,e^{-(2\pi (n+1)/a)\omega}\,e^{+i(\Delta/a)\omega}
 	\right].
 	\label{A10series}
 \end{align}
 Each integral is elementary:
 \[
 \int_{0}^{\infty}d\omega\,e^{-(p+i q)\omega}=\frac{1}{p+i q},\qquad p>0.
 \]
 Hence
 \begin{align}
 	G'(\Delta)
 	&=\frac{i}{4\pi a}\sum_{n=0}^{\infty}
 	\!\!\left[
 	-\frac{1}{\tfrac{2\pi n}{a}+i\tfrac{\Delta}{a}}
 	+\frac{1}{\tfrac{2\pi (n+1)}{a}-i\tfrac{\Delta}{a}}
 	\right]
 	=\frac{i}{4\pi}\sum_{n=0}^{\infty}
 	\!\!\left[
 	-\frac{1}{2\pi n+i\Delta}
 	+\frac{1}{2\pi (n+1)-i\Delta}
 	\right].
 	\label{A10sum}
 \end{align}
 
 By shifting the index in the second term $(m=n+1)$, this becomes a symmetric
 (bilateral) sum:
 \[
 G'(\Delta)=\frac{i}{4\pi}
 \left[
 -\sum_{n=0}^{\infty}\frac{1}{2\pi n+i\Delta}
 +\sum_{m=1}^{\infty}\frac{1}{2\pi m-i\Delta}
 \right].
 \]

  Relabeling $m\to n$ in the second sum and combining both series into a
  bilateral form gives
  \begin{align}
  	G'(\Delta)
  	&=\frac{i}{8\pi}
  	\sum_{n=-\infty}^{\infty}
  	\left(\frac{1}{2\pi n-i\Delta}-\frac{1}{2\pi n+i\Delta}\right).
  	\label{A10_symmetric}
  \end{align}
  
  \noindent
Now,
  \[
  \frac{1}{2\pi n-i\Delta}-\frac{1}{2\pi n+i\Delta}
  =\frac{2i\Delta}{(2\pi n)^2+\Delta^2}.
  \]
  Substituting into \eqref{A10_symmetric} gives
  \begin{align}
  	G'(\Delta)
  	&=\frac{i}{8\pi}\sum_{n=-\infty}^{\infty}\frac{2i\Delta}{(2\pi n)^2+\Delta^2}
  	=-\frac{1}{4\pi}\sum_{n=-\infty}^{\infty}\frac{\Delta}{(2\pi n)^2+\Delta^2}.
  	\label{A10_preIdentity}
  \end{align}
  
  \noindent
Now using
  the well-known series identity
  \[
  \sum_{n=-\infty}^{\infty}\frac{1}{n^2+x^2}
  =\frac{\pi}{x}\coth(\pi x)
  \]
  with $x=\Delta/2\pi$  gives
  \begin{align}
  	\sum_{n=-\infty}^{\infty}\frac{\Delta}{(2\pi n)^2+\Delta^2}
  	&=\frac{\Delta}{4\pi^2}\sum_{n=-\infty}^{\infty}
  	\frac{1}{n^2+(\Delta/2\pi)^2}
  	=\frac{1}{2}\coth\!\left(\frac{\Delta}{2}\right).
  	\label{A10_identity}
  \end{align}
  Inserting this into \eqref{A10_preIdentity} yields
  \begin{equation}
  	G'(\Delta)=-\frac{1}{8\pi}\coth\!\left(\frac{\Delta}{2}\right).
  	\label{A10_intermediate}
  \end{equation}

%

\subsection*{B}

\begin{eqnarray*}
	&& \frac{1}{2\pi} \int_{-\infty}^\infty e^{\pi \omega/2}\Gamma(-i\omega + \epsilon)(kV)^{i\omega -\epsilon}d\omega \nonumber\\
	&&= \frac{1}{2\pi} \int_{-\infty}^\infty e^{\pi \omega/2} \int_{0}^\infty t^{-i\omega + \epsilon -1} e^{-t}dt   (kV)^{i\omega -\epsilon}d\omega \nonumber\\
	&&= \frac{1}{2\pi} \int_{-\infty}^\infty e^{\pi \omega/2}  \int_{0}^\infty ( \frac{t}{kV})^{-i\omega + \epsilon -1} e^{-t}d(\frac{t}{kV})   d\omega \nonumber\\
\end{eqnarray*}
substituting $x = \frac{t}{kV}$ and assuming $V\neq 0$ gives
\begin{eqnarray*}
	&&= \frac{1}{2\pi} \int_{-\infty}^\infty  \int_{0}^\infty  e^{-i(log x+i\frac{\pi}{2}) \omega + \epsilon\log x   -\log x} e^{-kV x}dx   d\omega \nonumber\\
	&&=   i    \int_{0}^\infty \delta(\log x+i\frac{\pi}{2}) e^{ \epsilon \log x  -\log x} e^{-kV x}dx     \nonumber\\
	&&= e^{ikV}
\end{eqnarray*}

This only works if one assumes $V\neq 0$. If $V=0$, the equality ceases to exist.

 \subsubsection{Why $  [\int_{-\infty}^\infty \Gamma(-i\omega + \epsilon)(k'V)^{i\omega-\epsilon} e^{-ik'V}dV] = 0$}
 
 To see explicitly why the integral vanishes for $k' < 0$, one may extend $V$ to the complex plane and close the contour in the half-plane where $e^{-ik'V}$ decays, excluding a small indentation around $V=0$. The integrand $(k'V)^{\,i\omega-\epsilon}$ has only a branch point at the origin and no poles, so the closed-contour integral must be zero. The contribution from the large arc at infinity vanishes due to the exponential decay of $e^{-ik'V}$ in the chosen half-plane. 
 
 For the small semicircle around $V=0$, writing $V = r e^{i\theta}$ shows that the integrand carries the factor $(k'V)^{\,i\omega-\epsilon} \sim |k'|^{-\epsilon} r^{-\epsilon} e^{i\omega\ln r}$, and $dV \sim i r e^{i\theta} d\theta$. The magnitude of the integral is therefore proportional to $r^{1-\epsilon}$. Since $r^{i\omega}$ contributes only a phase, both upper and lower semicircle contributions vanish individually as $r \to 0$. Hence, with no enclosed singularity and vanishing boundary terms, the entire contour integral equals zero, implying that the inner $V$-integral is zero for $k' < 0$.
 
\section*{Data availability}
There is no data associated with the manuscript

\section*{Conflict of Interest}
 There are no conflict of interests
\section*{Acknowledgements}
We thank Dr. Rajeev Gupta for comments on some aspect of this  manuscript.

\end{document}